\documentclass[fleqn,10pt]{wlscirep}
\usepackage[utf8]{inputenc}
\usepackage[T1]{fontenc}
\usepackage{hyperref}
\usepackage{braket}
\usepackage{amsmath,amssymb,amsfonts}%
\usepackage{mathtools}
\usepackage{threeparttable}%
\usepackage{orcidlink}

\title{Quantum anomaly detection in real scarce data}

\author[1, *]{\orcidlink{0000-0003-3840-3609} Emanuele Casciaro} 

\author[2]{\orcidlink{0009-0008-6980-7095} Fabio Mascherpa} 

\author[2]{\orcidlink{0000-0002-7473-5945} Alfonso Amendola} 

\author[1]{\orcidlink{0000-0002-8366-4296} Filippo Caruso} 

\affil[1]{Department of Physics and Astrophysics, University of Florence, Via Sansone, 1, Sesto Fiorentino, 50019, Italy}

\affil[2]{DICOX/C – High Performance Computing Center of Excellence, DIT – Digital \& Information Technology, Eni S.p.A., Via Emilia 1, San Donato Milanese, 20097, Italy}

\affil[*]{emanuele.casciaro@unifi.it}

\begin{abstract}
Anomaly detection on small and unbalanced datasets remains very challenging in machine learning, although this scenario is common in several domains, including healthcare, cybersecurity, finance, and energy. Data augmentation and generative AI may mitigate training-data scarcity, but they often fall short because anomalies are, by definition, unpredictable, rare, and highly diverse events compared to high-probability normal data. Overfitting to pseudo-anomalies, model collapse, high-dimensional data, uninterpretable black-box models, and validation challenges are typical issues limiting their practical applicability. In this context, quantum machine learning may provide a promising and more sustainable avenue because it can enable more interpretable models with far fewer trainable parameters and smaller datasets, implementable on energy-efficient quantum hardware. Here, we propose a novel two-step hybrid classical--quantum architecture for sequential data and test it on a realistic scenario in the global energy-transition domain, i.e., automated anomaly detection in large-scale photovoltaic plants. The achieved generalization capability and competitive prediction accuracy may pave the way for new hybrid learning models able to exploit the continuously increasing power of cloud-available and more sustainable quantum accelerators integrated with more traditional energy-hungry High Performance Computing resources.
\end{abstract}
\begin{document}

\flushbottom
\maketitle
%
%
\thispagestyle{empty}

\section*{Introduction}
In the context of learning theory, anomaly detection (AD) addresses the problem of identifying whether a sample (or a time window) is \textit{normal} or \textit{anomalous}, whereas anomaly classification (AC) aims to assign an anomalous sample to a specific fault type. An anomaly is commonly defined as an observation that does not conform to the expected behavior of the system~\cite{chalapathy_deep_2019}. In real-world settings, AD/AC datasets are therefore highly unbalanced (anomalies are rare by definition), and crucially, anomaly labels are often scarce or expensive to obtain (i.e., requiring expert inspection), which can severely limit solutions based on labeled data.
As the reliability of digital systems becomes a critical aspect, AD has become increasingly relevant in recent research, with notable applications in fields ranging from cybersecurity~\cite{ten_anomaly_2011, hong_integrated_2014} to financial fraud detection~\cite{ahmed_survey_2016}, IoT systems~\cite{said_efficient_2021}, smart cities~\cite{garcia-font_comparative_2016} and energy applications~\cite{perez2025recent}, to name just a few.
In industrial contexts, AD and AC systems find valuable use in predictive maintenance, one common application being in large-scale solar photovoltaic plants~\cite{zulfauzi_anomaly_2023}. Predictive maintenance aims to prevent or reduce plant downtime due to major malfunctions by repairing or replacing defective components at the first signs of divergence from their expected behavior. Such anomalies in component operation may be undetectable or too costly to track through human inspection but easily recognized by sensors or imaging. This can be exploited by devising software capable of recognizing unexpected behavior from real-time sensor data or automatically inspecting panels for visible and non-visible defects~\cite{vlaminck_region-based_2022} (i.e., by thermal imaging) and sending warnings to prompt a timely response when needed.

A more complex formulation often related to the problem of predictive maintenance involves dealing with sequential data, often provided by a large number of sensors. The time dimension present in the so-called time series anomaly detection (TS-AD) introduces additional challenges due to the dynamic nature of the system, which makes it more difficult to reliably define and consequently identify an anomaly. In TS-AD, unexpected behavior can be classified into three main categories~\cite{sorbo_navigating_2024}. Point anomalies, or outliers, are the simplest category. These are individual data points that are inherently anomalous, in the sense that they do not belong to the general data distribution. Contextual anomalies are more difficult to identify; a data point that would not be anomalous \textit{per se} becomes so in relation to its specific neighborhood or context. Finally, collective anomalies are anomalies that concern a group of data points that are close to each other and collectively deviate from the underlying distribution. In most TS-AD applications, it is more interesting to capture the last two types of anomalies, which can only be identified by studying the correlations emerging among data points in the sequence over time.

In recent years, many machine learning algorithms have been proposed to solve the AD problem, leveraging assumptions about the data distribution to identify outliers. The isolation forest algorithm~\cite{iforest}, for instance, flags as anomalies the most isolated sampled points and is used in the predictive maintenance of IoT-aided PV systems~\cite{singh_thakur_efficient_2023}. Other methods flag as anomalies those instances that fail to associate with any other point in the dataset~\cite{kwon_survey_2019}, or learn a decision boundary based on previously seen sets comprising only non-anomalous data. The One-Class Support Vector Machine (SVM)~\cite{manevitz_one-class_2002}, for example, adapts the SVM formulation to the AD task and is widely used in monitoring large industrial plants~\cite{ferreira_using_2022}. However, all these methods are only capable of dealing with point anomalies and are not suitable for processing temporal information; in the context of TS-AD, the Autoregressive Integrated Moving Average (ARIMA) family of models~\cite{moschini_anomaly_2021} allows processing sequential information and is commonly used to monitor individual critical components (e.g., a turbogenerator~\cite{kozitsin_online_2021}). Basic ARIMA models can only handle univariate sequences, but the related and more general Vector Autoregression family of models~\cite{zivot_vector_2003} enables working with sequences of vectors as well, extending the scope of the methods to more realistic settings, including those with complex, multivariate data.

Recent advances in ML have led to the development of deep learning methods, which employ large parameterized models to learn a latent representation of the data and subsequently use this information to learn the desired target function. A widely used pattern in standard unsupervised AD is the autoencoder architecture~\cite{sakurada_anomaly_2014}, which learns to reconstruct non-anomalous data from a lower-dimensional representation and flags as anomalous any data point whose reconstruction error exceeds a certain threshold. This approach assumes that the dataset can be described by a set of latent variables, which are learned and used to fit a training set containing only sane data due to the scarcity of anomalous samples. Other techniques build upon a similar concept to exploit the scarcity of labeled anomalies; in~\cite{hojjati_dasvdd_2024}, a deep learning architecture is used to learn a latent representation of a dataset alongside a minimal hypersphere $\mathcal{S}$ that contains all points, signaling at test time every point projected by the model that does not belong to $\mathcal{S}$ as an anomaly, with promising results in industrial AD benchmarks on real-time production images~\cite{huang_improved_2025}.

Over the years, several deep learning architectures have been proposed to deal with correlations within data sequences; the Convolutional Neural Network (CNN)~\cite{kiranyaz_1-d_2019} method learns several one-dimensional convolutional filters from the data, allowing the extraction of some information from the sequence, at the cost of fixing the size of the learned sequence; this lightweight model can be successfully applied in the monitoring of simple, non-interconnected systems with short-term memory, as in~\cite{apeiranthitis_predictive_2024}, where a convolutional model is used in the predictive maintenance of mechanical equipment, using the data relative to their vibrations. The Recurrent Neural Network (RNN)~\cite{schmidt_recurrent_2019} method, which is frequently used in AD in PV plants~\cite{yi_anomaly_2025}, improves this aspect by storing a hidden state vector $h$, which is updated every time a new element is processed, allowing the encoding of sequences with indefinite length, at the cost of an exponentially vanishing gradient as the sequence grows, due to the repeated operations on the same weight matrix. The Long Short-Term Memory (LSTM) cell~\cite{hochreiter_long_1997} enables a partial update of the recurrent cell, leading to the gradients vanishing more slowly and therefore larger treatable datasets; one of the most recent and popular solutions is the Transformer architecture~\cite{vaswani_attention_2017}, which encodes every element of a sequence with arbitrary length using a non-linear mapping of the “query, key and value" representation, and then computes a final embedding from them, taking into account every other element and its intersections within the sequence. This method is successfully demonstrated in~\cite{wirawan_photovoltaic_2024} as a means to reliably isolate occurrences of anomalous energy output of PV systems caused by harsh climate conditions.

A common trend is the adoption of increasingly larger models as the size of the problem grows. Although this reflects the increasing complexity of the task at hand, this choice is not always optimal due to the new set of challenges that may arise when dealing with a high number of trainable parameters. These include a more expensive training procedure that requires more time and more data, and less predictable, more black-box behavior as the model’s size and complexity grow.

Quantum Machine Learning (QML) may offer a way around the numerical challenges presented by such problems by taking advantage of the unique characteristics of quantum algorithms to bypass some of the computational limits of classical ML methods~\cite{schuld_introduction_2015}, including in the field of AD~\cite{kyriienko_unsupervised_2022}. In particular, Quantum Neural Networks (QNN) have attracted much interest in the past few years. Similarly to its classical counterpart, a QNN uses a set of parameters to learn a quantum function, which is capable of expressing complex relationships not possible in classical ML by exploiting superposition and entanglement to express non-classical states~\cite{dalzell_quantum_2025}.

In this work, we try to harness the power of quantum computation to perform anomaly classification in a realistic energy transition problem, assessing the capabilities of QNNs to achieve desirable performance using smaller (in terms of parameters) models, leading to more efficient and less expensive training. The article is structured as follows: in Section~\ref{sec:results}, we showcase and discuss the results of a test application of our method over a realistic dataset, and in Section~\ref{sec:methods}, the theoretical and empirical background of our work is provided, followed by the description of our hybrid training procedure. 

\section*{Results}\label{sec:results}
\begin{figure}
    \centering
    \includegraphics[width=\linewidth]{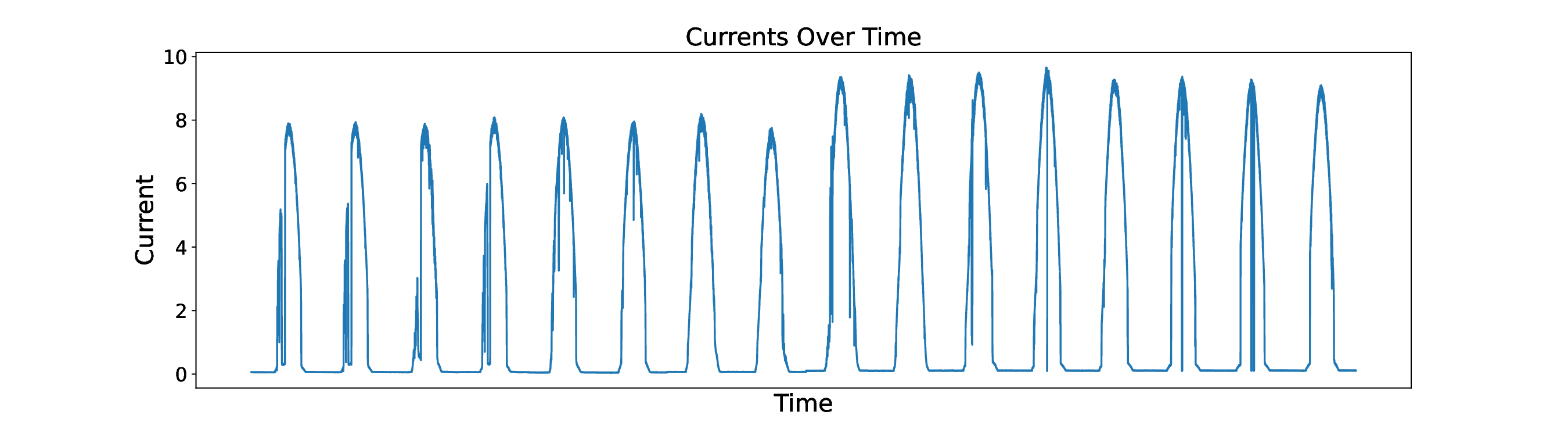}
    \caption{Current level of the first string of panels sampled over the course of the experiments. The peaks represent daytime activity, while the inactive regions are associated with nighttime.}
    \label{fig:day}
\end{figure}

Given that assessing the effectiveness of the models in realistic industrial settings is crucial in the development of AD techniques, the authors in \cite{panels_dataset} synthetically generated a dataset describing the simulation of a real solar panel power plant, inducing different kinds of faults during the simulation period, with the goal of matching a real-world scenario of deployment. In order to provide data as close as possible to a real use case, the simulation was performed using real instruments, and the errors introduced were derived from an artificially induced hardware failure. The simulation spanned a period of 16 days (Figure~\ref{fig:day}) with a sampling rate of 1 Hz, providing data gathered from two different strings of solar panels.

\begin{figure}
    \centering
    \includegraphics[width=\linewidth]{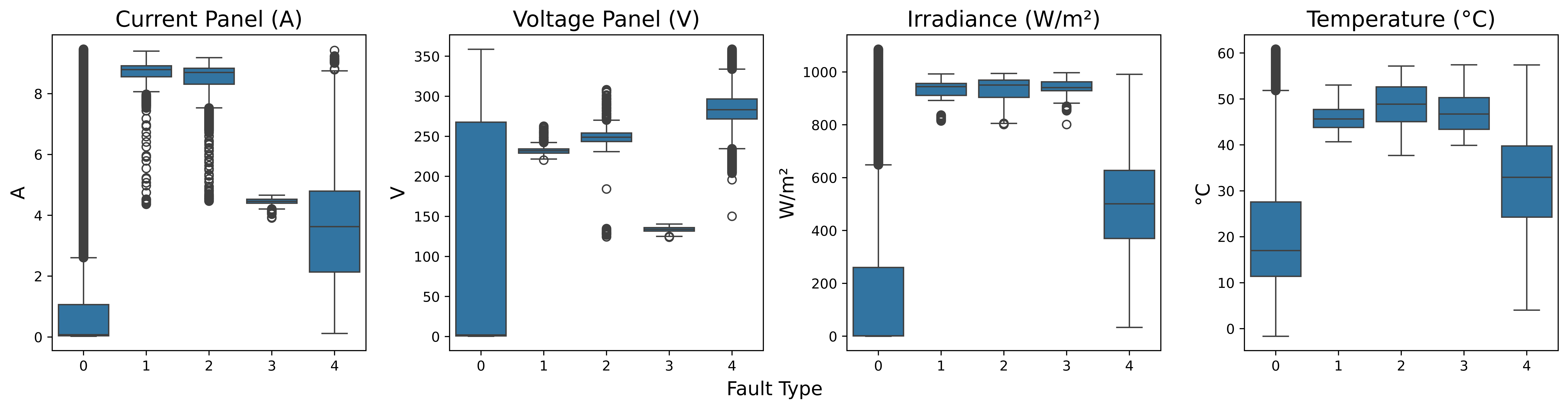}
    \caption{Per-class feature distribution: due to the sane data being distributed all over the feature space, it is difficult to correctly identify the different types of fault instances. }
    \label{fig:boxplot}
\end{figure}

\begin{figure}
    \centering
    \includegraphics[width=\linewidth]{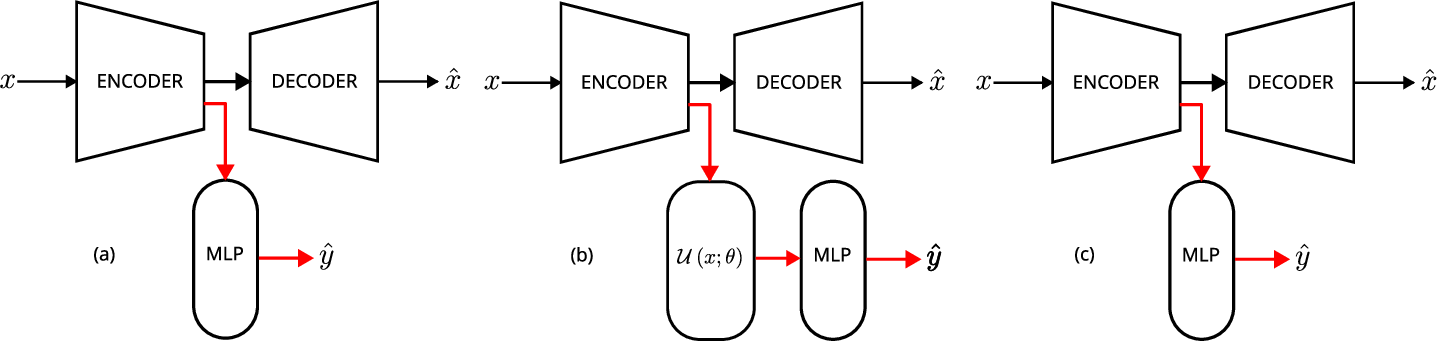}
    \caption{Different hybrid interactions analyzed: no interactions (left) is compared against a hybrid model where the quantum layer is employed only at the classification level (ours, center), or from the beginning of the training (right). The black and red lines represent the flow of the input during the first and second training phase.}
    \label{fig:arch-ablation}
\end{figure}

Each data point describes the system status with the voltage and current values of both the strings of panels, alongside the irradiance and temperature of the system. Artificially introducing the anomalies inside the environment provides an automatic labeling of the data, which represents the kind of anomaly induced, which can be classified as the presence of a \emph{short circuit} (label $1$), \emph{degraded} (label $2$), \emph{open circuit} (label $3$) or \emph{shadowing} (label $4$), with non-faulty instances, which make up for the majority of the dataset (Figure~\ref{fig:boxplot}), represented with label $0$. 
The models are trained to minimize the \textit{weighted cross-entropy} loss, where each class has weight $w_i = 1 - r_i$, and $r_i$ is the ratio of occurrence of the class reported in Table~\ref{tab:class-distr}.
The standard metric for evaluating classification models is the \emph{accuracy}, defined, given the \textit{true} and \textit{false} \textit{positive} and \textit{negative} prediction outputs $TP$, $FP$, $TN$ and $FN$, as 
\begin{center}
$$Acc = \frac{TP+TN}{TP+FN+FP+TN},  $$   
\end{center}
which measures the fraction of correct predictions over all the data. However, assessing the performance of an AD model is not a trivial task: using naive metrics such as accuracy may lead to misleading information, since it does not take into account the unbalanced composition of the dataset, usually consisting of few anomalies and an abundance of regular data points, encouraging models to always flag points as non-anomalous, achieving high accuracy, while being clearly under-performing for the desired task. A more encouraged metric is the\begin{center}
$$F_1=\frac{2TP}{2TP+FN+FP}$$
\end{center} 
score, which gives the same importance to both $precision=\frac{TP}{TP+FP}$ and $recall=\frac{TP}{TP+FN}$, rewarding different behaviors in classification. Although this metric takes into consideration the imbalance of the data, it does not depend on the number of \emph{true negatives}, representing, even if to a reduced magnitude, a different bias in the evaluation. A more comprehensive metric is the \emph{balanced accuracy}, which is the mean of the \emph{true positive rate} and \emph{true negative rate}, which indicates the probability of the model classifying a sample as each class given the ground truth.
\begin{center}
$$Acc_b \vcentcolon = \frac{1}{2} \left( \frac{TP}{TP + FN} + \frac{TN}{TN + FP} \right)$$
\end{center}
The problem, expressed in both the anomaly \textit{detection} and \textit{classification} formulation,
is used to benchmark the two-step training architecture described in Section~\ref{sec:methods}, which first learns a more compact representation of the data through a classical network and then uses a hybrid classifier, which shares components with the compression network, to perform prediction over the learned representation. The model performances are compared with two different architectures.
The first is a fully classical architecture, constructed following~\cite{hae} by replacing the quantum layer of the hybrid classifier with a fully connected layer with equivalent latent dimension size $N=2^Q$, with $Q$ being the size of the latent dimension of the learned representation and, consequently, the number of qubits used. Since the reconstruction objective function used for learning the representation is unsupervised, it is cheap to train and could potentially be considered fixed once trained with enough data. Achieving good performance with a quantum model implies the possibility of training a reliable supervised AD and AC model using few training samples, with a lower risk of overfitting compared to a classical-only solution. The other model follows~\cite{hae} by using the QNN layer as an additional layer in the compression network, shifting the training of the quantum layer from the classification-only objective to also include the reconstruction task. At the cost of a more expensive initial training due to the presence of the quantum layer, the classification training becomes more straightforward, with only a small classification head to train.
These variations, summarized in Figure~\ref{fig:arch-ablation}, are done with the goal of assessing whether the quantum layers actually benefit the hybrid networks, and analyzing their contribution according to their position inside the architecture.

\begin{figure}
\centering
            \includegraphics[width=\linewidth]{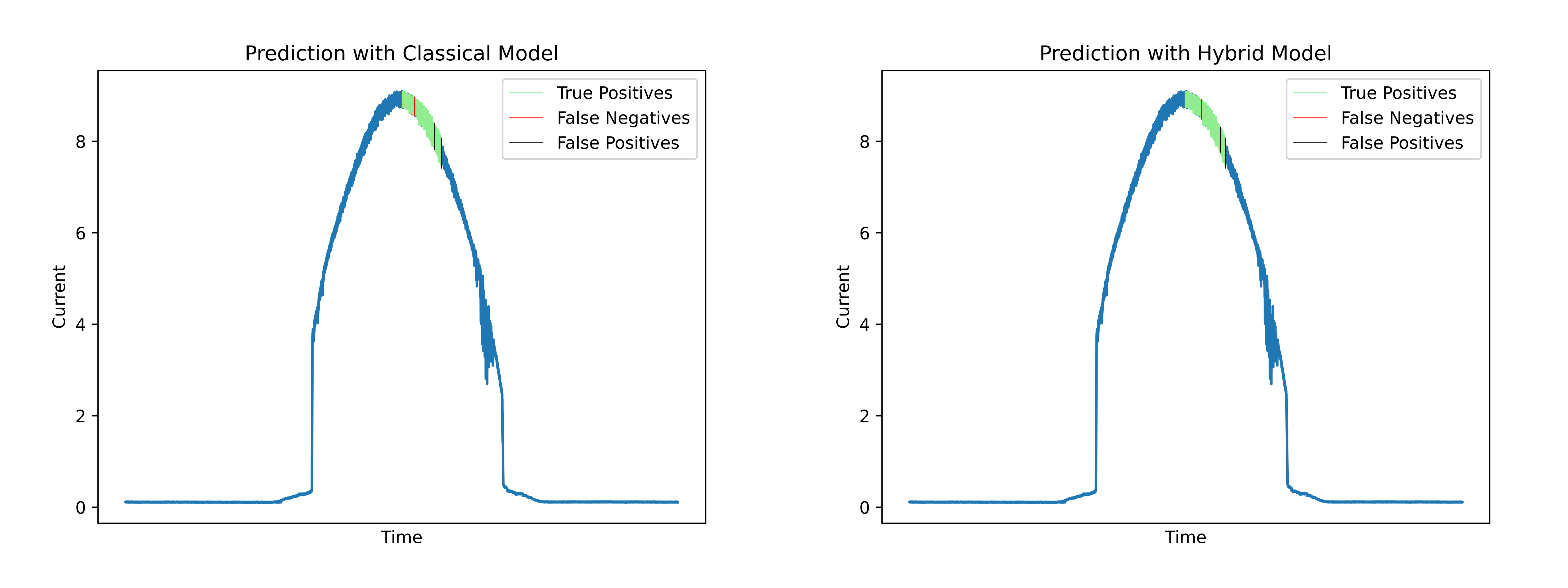}
    \caption{Anomaly detection over the last day of sampling. Both classic (left) and hybrid model (right) perform well, with the hybrid model avoiding some misclassification at the start of the sequence.}\label{fig:binary_time_plot}
    
    \end{figure}

In the AD task, whose results are reported in Table~\ref{tab:binary_results}, both the classical solution and our hybrid implementations perform similarly well, achieving near-perfect scores in all metrics, resulting in accurate predictions reported in Figure~\ref{fig:binary_time_plot}. 

While the scores for our scheme and the fully classical case are similar, our hybrid classification layer employs far fewer parameters, reducing their number by a factor of $\sim 40$, lowering both the risk of overfitting and the training cost. On the other hand, the application of the quantum layer at the encoder level breaks the autoencoder structure, resulting in a similarly small number of parameters but much poorer predictions. It is also worth noting that following the construction of the classical classifier described above, the parameter gain from the classical to the hybrid model would actually be much greater; however, since both classical and hybrid models achieve near-perfect scores, the key question to rank their performance becomes which model can solve the problem with fewer parameters and hence lower training costs. In this specific instance, the size of the classical classifier can be reduced from $\approx 267k$ to $\approx 4.4k$ by exploiting the structure of this problem without affecting the quality of the solutions. Therefore, in more complex tasks, assuming both classical and hybrid models retain comparable performances, we could have observed an exponential gain in parameter efficiency, up to a factor of $\approx 2.4k$.

In the AC case, our hybrid model manages to achieve better test-time scores than the classical model, as reported in Table~\ref{tab:scores}. Similarly to the AD case, it again outperforms the hybrid autoencoder--based approach, suggesting that quantum layers may yield different outcomes when placed in different positions in a hybrid neural network.
In the per-class error breakdown reported in Table~\ref{tab:per_class_scores}, we can observe that the hybrid model is less prone to incorrectly classifying sane data as anomalies, resulting in an overall reduction in false negatives. Even in the distinction between sane data (class $0$) and data associated with the \emph{partial obscuration} anomaly (class $4$), which is hard to correctly identify, the hybrid model outperforms the classical counterpart.
Similarly to the AD case, even though the hybrid model generally outperforms the classic model, the key difference in this case is the number of parameters employed in the classification head of the network: while the classic model requires $\mathcal{O}\left(2^q\right)$ parameters, the hybrid model achieves comparable inference capabilities by requiring only $\mathcal{O}\left(q\right)$ learnable quantum gates: this exponential reduction in the number of weights can result in a faster fine-tuning of the network and a lower risk of overfitting the training data, leading to more reliable models. It is notable that the number of quantum parameters required in both AD and AC cases is the same, the only difference being the size of the final layer adapting to the number of possible classes.

\begin{figure}
    \centering
    \includegraphics[width=0.6\linewidth]{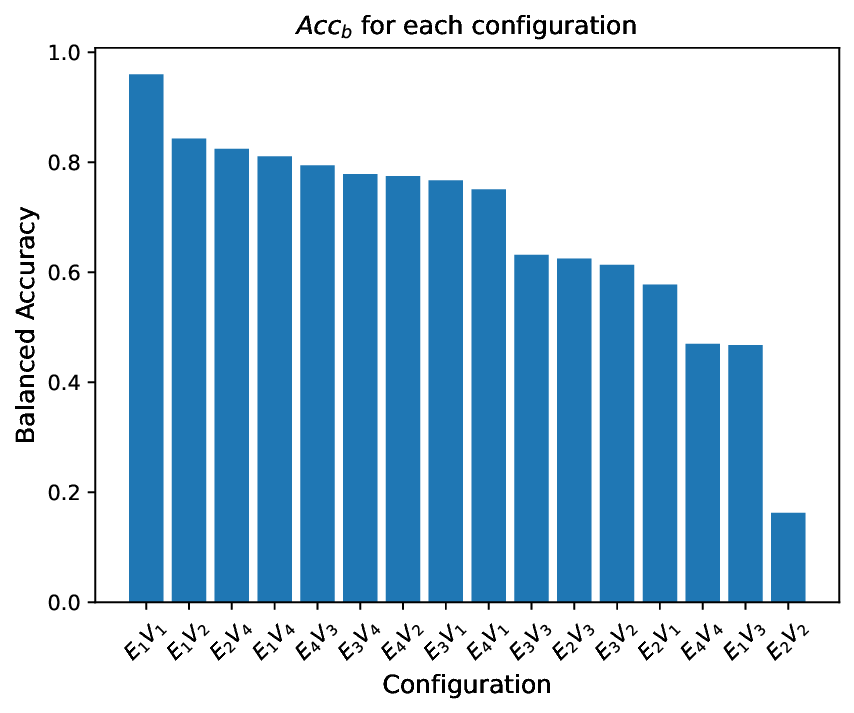}
    \caption{$Acc_b$ of each configuration of ansatz construction: without changing the measurement of the circuit (in our case $\braket{Z_i}$), the family of states $\ket{\psi(x;\theta)}$ associated with the circuit can have a great impact on the output of the layer, affecting the classification performances.}
    \label{fig:ansatz-ablation}
\end{figure}

To assess the relevance and effectiveness of different circuit constructions, we tested our architecture using different ansätze, which are described in Table~\ref{tab:ansatz-modules}.
From the results, summarized in Figure~\ref{fig:ansatz-ablation}, it is possible to observe that circuit construction is a major source of variability in classification performance: a simple circuit, given enough re-uploading blocks, outperforms more sophisticated circuits while keeping the number of required operations small, reducing noise in real-hardware applications and making the method suitable even for testing on currently available noisy intermediate-scale quantum (NISQ) devices.

\section*{Methods}\label{sec:methods}

\begin{figure}
    \centering
    \includegraphics[width=\linewidth]{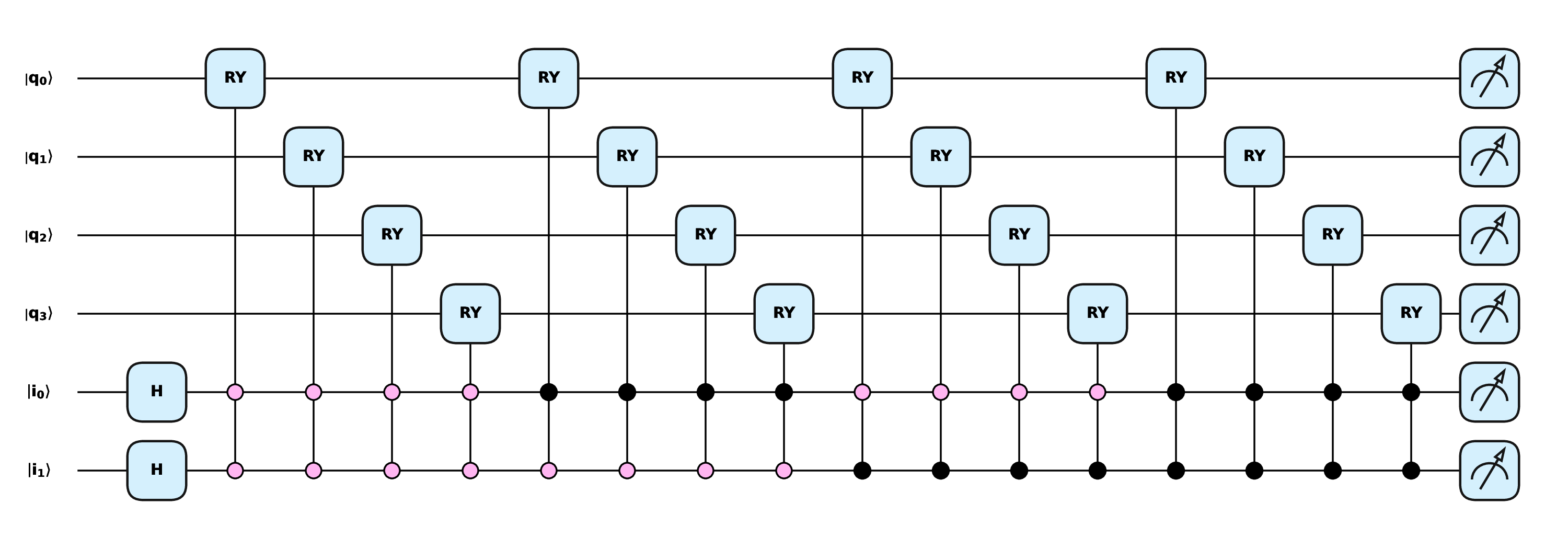}
    \caption{Circuit representation of the feature map encoding a sequence of $d=3$ dimensional data of length $T=4$. Using different embedding strategies is possible to reduce the amount of qubits required for each element of the sequence, at the expense of a deeper circuit. }
    \label{fig:seq-embedding}
\end{figure}

A generic quantum algorithm applies a series of operations $\mathcal{U}_i$ to an initial reference state $\ket{0}$, evolving it into the final state $\ket{\psi}=\left(\prod_i \mathcal{U}_i\right) \ket{0}$, which is then measured to retrieve a classical value corresponding to the result of interest. In QNNs, the input data is encoded using a \textit{feature map}, a parameterized operator that transforms the initial state into $\ket{x}=\mathcal{U}_{\Phi(x)}\ket{0}$, and then processed using a trainable parameterized operator $V_\theta$, which adds degrees of freedom to the learned quantum function. Similarly to how it is possible to stack more \textit{perceptrons} into a \textit{multi-layer perceptron}~\cite{mlp_def}, which adds more expressive power to the network, one can repeat the feature map and variational layer, obtaining the final circuit $\mathcal{U}(x;\theta)=\prod_i V_{\theta_i}\mathcal{U}_{\Phi(x)}$, creating the \textit{data-reuploading scheme}~\cite{reupload_P_rez_Salinas_2020}. Most implementations use, as output of the model, the expectation value of an observable $\mathcal{O}$, resulting in the output value
\[
y = \braket{0|\mathcal{U}(\theta; x)^\dagger\mathcal{O}\,\mathcal{U}(\theta; x)|0}.
\]
This can then be used to compute a \textit{loss} function, which evaluates the model and enables optimization of the parameters through classical algorithms, i.e., the \textit{gradient descent} update rule. This scheme, similar to the classical case, has been successfully applied in different contexts~\cite{tacchino_quantum_2020, cong_quantum_2019, li_quantum_2023}, showing promising results in several benchmarking problems. However, these methods are fairly limited in problem size due to the hardware required for handling larger problems, making them ill-suited for real-world tasks. Another limitation of QNNs is their ability to handle sequential information: feature maps capable of expressing positional information require supplementary qubits (Figure~\ref{fig:seq-embedding}), making it impossible to operate with arbitrarily long sequences.
The \textit{Quantum RNN} model~\cite{li_quantum_2023} tries to solve the problem by storing the state information in a fixed-size register every time a different time-step is encoded; however, for larger sequences the circuit becomes too deep, requiring mechanisms for qubit resets and introducing additional operational overhead, which hinders performance. In the \textit{Quantum LSTM}~\cite{chen_quantum_2022, zhou_novel_2024} the state vector is classical, and different quantum circuits are used to decide the update of the cell, eliminating the scaling issue at the cost of reducing the contribution of quantum effects to the state evolution. A suitable solution is to rely on \textit{hybrid} models that can combine the adaptability of classical architecture for processing sequential information into a fixed-size embedding, used by the quantum circuit, that processes it in a more complex state space. Hybrid models have applications in classification~\cite{arthur_hybrid_2022, senokosov_quantum_2024}, anomaly detection~\cite{wang_quantum-classical_2023, hae}, and forecasting~\cite{schetakis_data_2025} with competitive results compared to standard NN architectures, while also being able to reduce the number of qubits required for processing the data, and hence the amount of resources required.

Inspired by~\cite{hae}, where the authors used a hybrid network to compress the data, which is then used to perform unsupervised AD using the \textit{isolation forest}~\cite{iforest} algorithm, we employ a transformer-based \textit{autoencoder} to learn a more compact representation of the sequence, made of an \textit{encoder} $E: \mathbb{R}^d\rightarrow\mathbb{R}^p$, which compresses the sequence, and a decoder $D:\mathbb{R}^p\rightarrow\mathbb{R}^d$ that recreates the original input and is trained over the \textit{mean squared error} of the data reconstruction \[\mathcal{L}(E, D, \vec{x_t})=\sum_{t=1}^T\left\|x_t - D\left(E(\vec{x_t})_t\right)\right\|^2.\]

The autoencoder is based on the Transformer architecture, which uses the self-attention mechanism to compute an embedding for each element of the sequence based on its relationship with the other components: the attention mechanism is masked using a triangular matrix, in order to avoid non-causal interactions. The encoder is composed of several stacked blocks, which reduce the dimension of the data at each step, and the decoder retains a symmetric structure, restoring the initial dimensionality.

Once the encoder learns a suitable representation for the reconstruction task, the decoder is discarded, and the encoder is fine-tuned alongside a hybrid classifier $H$, comprised of a QNN layer alongside a shallow classical prediction. The classification training minimizes the \textit{weighted cross-entropy loss} \[\mathcal{L}(E, H, \left\{x_t\right\}_{t=1}^T, y) =-\sum_{k=1}^{K} \lambda_k y^{(k)} \log \left[H\left(Agg\left(E\left(\left\{x_t\right\}_{t=1}^T\right)\right) \right)\right]^{(k)},\]
with $\lambda_k>0$ the weight of each class, used to counter the class imbalance, and $Agg: \mathbb{R}^{T\times p}\rightarrow \mathbb{R}^{p}$ an aggregation function: different aggregators have been tested, including uniform and weighted mean $Agg\left(\left\{x_t\right\}_t\right)=\sum_t w_t x_t$, with $w_t=\frac{1}{T}$ in the case of uniform mean, otherwise a learnable parameter; however, the selected aggregator is a last-element selector $Agg\left(\left\{x_t\right\}_t\right)=x_T$, due to the similar classification outcome and lightweight cost.

The classifier is a hybrid network comprising a trainable quantum circuit and a classical prediction head, which helps the quantum classifier adapt the output to the problem specifics without post-network processing and aids in abstracting the design of the measurement process from the dimensionality of the problem, delegating the task of reconstructing the right output from the measurement to a classical processor.

\begin{figure}
    \centering
    \includegraphics[width=0.7\linewidth]{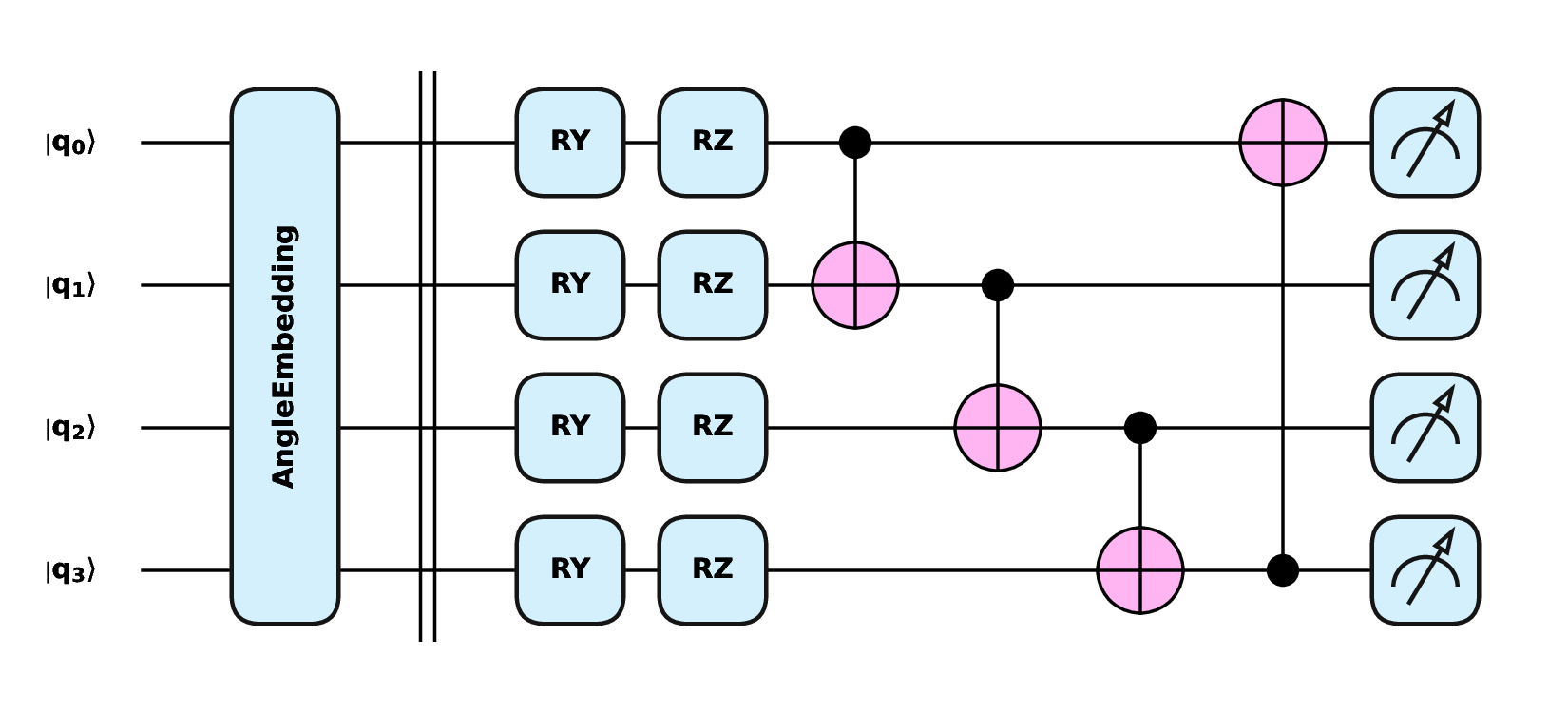}
    \caption{Ansatz used during the training: the data is encoded through the \textit{angle embedding} $\ket{x}=\otimes_{i} R_X(x_i)\ket{0}$ and processed using a variational layer.}
    \label{fig:ansatz}
\end{figure}

The quantum circuit, as depicted in Figure~\ref{fig:ansatz}, is a QNN layer comprised of a feature map which encodes the data using the \textit{angle embedding} scheme over the \textit{Pauli X} axis, resulting in the state \[ \ket{x}=\otimes_k \left(\cos \frac{x_k}{2}\ket{0}-i \sin \frac{x_k}{2}\ket{1}\right),\]
where $\otimes_k$ denotes the tensor product of all qubits, labeled by $k$.
The quantum state evolves using a set of single-qubit rotations over the Pauli axes $Y$ and $X$, with a set of controlled-$X$ gates applied in a ring connectivity, which introduces entanglement in the circuit. The whole scheme is repeated $R$ times according to the data-reuploading scheme, and at the end of the circuit the $\braket{Z}$ expectation value of each qubit is measured, ensuring an output $o$ bounded in norm by $\|o\|_\infty\leq 1$. The prediction head is a \textit{linear probe}~\cite{alain_understanding_2018}, with the sole purpose of computing the logit vector from the expectations of the qubits, relaxing the circuit design from the constraint $\#classes \leq \#qubits$, using the most information about the output state.

\begin{figure}
    \centering
    \includegraphics[width=.7\linewidth]{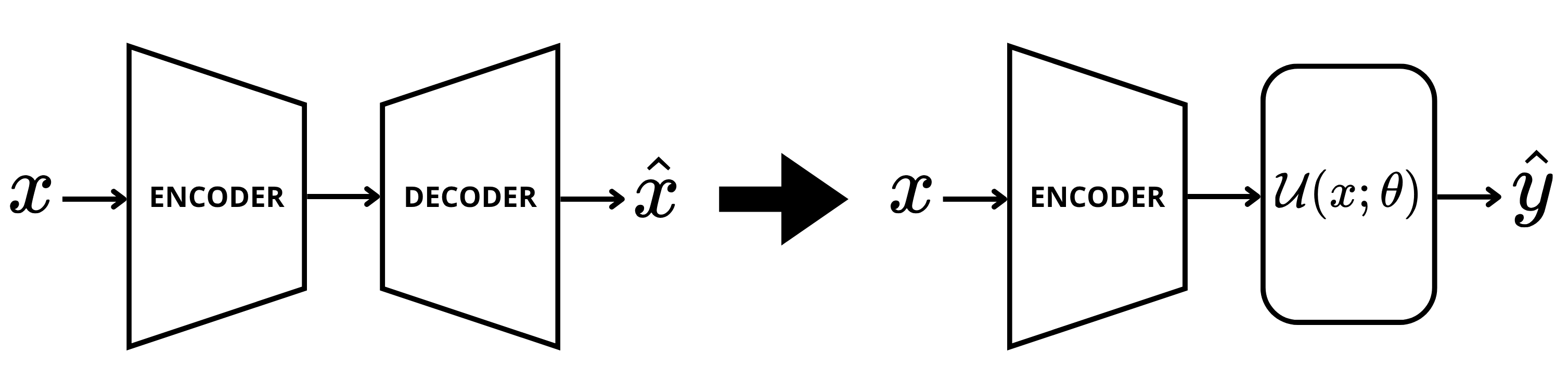}
    \caption{Training scheme of the proposed architecture: once the autoencoder learns a suitable representation of the data, the decoder is replaced by the hybrid classifier.}
    \label{fig:training}
\end{figure}

By using this two-step training procedure, summarized in Figure~\ref{fig:training}, it is possible to work efficiently with sequences using hybrid networks. Once we are able to extract a sufficient amount of fixed-size information from the sequence, we can use the quantum classifier to process it in a more expressive computational space, even in the case of real-world complex problems.

\subsection*{Implementation Details}
Both hybrid and classical solutions are implemented using PyTorch~\cite{paszke_pytorch_2019}, with the aid of Pennylane~\cite{bergholm2022pennylaneautomaticdifferentiationhybrid}, used in the realization and training of the quantum circuits. All the solutions tested were trained for 10 epochs with the reconstruction task and 15 epochs over the classification objective. Training, validation and test split are obtained through sub-sampling of the whole dataset, ensuring to evenly distribute the anomalies, concentrated towards the end of the dataset, across all the splits.

\begin{table}[b]
    \centering
    \caption{Per-class distribution of each type of anomaly. The \textit{Sane} state makes up for almost all the dataset size.}
    \begin{tabular}{c c c c c c}
    \toprule
        \multicolumn{6}{c}{Data incidence (\%)} \\
      \midrule
         Sane &  S. Circuit & Degradation & O. Circuit & Obscuration & All Anomalies \\
         0.847 & 0.004 & 0.008 & 0.004 & 0.137 & 0.153\\
    \bottomrule
    \end{tabular}
    \label{tab:class-distr}
\end{table}

\begin{table*}[b]
                \centering
               \caption{Experiment results for the binary prediction task. Our proposed model and its classical counterpart perform similarly well, while the HAE model fails to reliably detect anomalies.}
               \label{tab:binary_results}
               \begin{threeparttable}
                   
                \begin{tabular}{ccccc}
                    \toprule
                    Model & $Acc$ & $Acc_b$ & $F1$ & \emph{\#Parameters} \\
                    \midrule
                    
                    AE + QNN Classifier (ours) & 0.9952 & \textbf{0.989} & 0.9955 & 82\\
   Hybrid AE + Classical Classifier & 0.7796 & 0.5 & 0.8617 & 18\\
       AE + Classical Classifier & \textbf{0.9956} & 0.9877 & \textbf{0.9959} & $\approx4400$\tnote{1}\\
                    \bottomrule
   \end{tabular}
  \begin{tablenotes}
      \item[1] Despite using fewer parameters than required by the model construction, the final score is not affected. 
  \end{tablenotes} 
  \end{threeparttable}
\end{table*}
\begin{table}[b]
    \centering
        \caption{ 
        Classification scores in the AC task of the proposed models: while the HAE model has fewer parameters in the classifier with respect to our model, it shares a similar parametrized structure in the encoder. 
        }
    \begin{tabular}{c c c c c}
        \toprule
        Model & $Acc$ &  $Acc_b$ & $F1$ & \emph{\#Parameters} \\
        \midrule
        AE + QNN Classifier (ours) & \textbf{0.9357} & \textbf{0.9761} & \textbf{0.9418} & 109 \\
        Hybrid AE + Classical Classifier & 0.8515 & 0.9293 & 0.8789 & 45 \\
        AE + Classical Classifier & 0.8784 & 0.9549 & 0.9005 & $\approx4400$ \\
        \bottomrule
        \end{tabular}
        \label{tab:scores}
\end{table}
\begin{table}[b]
    \caption{Per-class accuracy score of the best performing models: our solution is less prone to flagging sane instances as anomalous data, resulting in a low false-positive count.
   \centering
    \label{tab:per_class_scores}}
   \begin{tabular}{c c c c c c}
    \toprule
    Model & Sane &  S. Circuit & Degradation & O. Circuit & Obscuration \\
    \midrule
    AE + QNN Classifier (ours) & \textbf{0.9481} &  \textbf{0.9967} & 0.974 & \textbf{1} & 0.9653 \\
    
    AE + Classical Classifier & 0.8881 & \textbf{0.9967} & \textbf{0.9846} &  \textbf{1} & \textbf{0.9734} \\
    \bottomrule
    \end{tabular}
\end{table}

\begin{table}[b]
\hfill
\begin{minipage}{0.4\textwidth}
\centering
    \begin{tabular}{l cc}
        \multicolumn{3}{c}{Encoding Layer}\\
        \toprule
        \#&Strategy&Entanglement\\
        \midrule
        1&$AE_X(x)$& No entanglement\\
        2&$AE_X(x^3)AE_X(x^2)AE_X(x)$&No entanglement\\
        3&Controlled $AE_X(x)$& Ring connectivity\\
        4&Controlled $AE_{\vec{\sigma}}(x)$& Ring connectivity\\
        \bottomrule
    \end{tabular}
\end{minipage}%
\hfill
\begin{minipage}{0.4\textwidth}
\centering
    \begin{tabular}{l cc}
        \multicolumn{3}{c}{Variational Layer}\\
        \toprule
        \#&Strategy&Entanglement\\
        \midrule
        1&$R_XR_Y$& CX Ring\\
        2&$R_XR_YH$& CX Ring\\
        3&$R_ZR_XR_YH$& CX Ring\\
        4&$R_{\vec{\sigma}}$& No entanglement\\
        \bottomrule
    \end{tabular}
\end{minipage}%
\hfill
    \caption{Encoding and variational strategies used in the construction of the ansatz; each encoding is tested in combination with every variational layer proposed. $AE_{a}(\cdot)$ denotes the \textit{angle encoding} with respect to the axis $a$, with $\vec{\sigma}$ indicating a general Pauli axis.}
    \label{tab:ansatz-modules}
\end{table}

\clearpage

\section*{Discussion}
We presented an application of hybrid QNNs in the context of anomaly detection and classification, targeting the reliability of PV systems, showing that hybrid techniques can achieve competitive performance compared to classical models while using far (sometimes even exponentially) fewer parameters. To effectively use QNNs for AD on sequential data, we proposed a two-step training method that couples a QNN with an autoencoder, which is initially trained to learn a more compact and suitable representation of the sequence. This representation can then be used by the quantum layer to obtain the final prediction. By applying this method to a dataset mimicking real-world solar-plant faults, we achieve better performance than similar hybrid algorithms, with prediction accuracies comparable to classical approaches but significantly greater parameter efficiency in the classification head. This improvement carries over to the AC task, enabling the classification of highly unbalanced data using few parameters and encouraging better inference-time generalization capabilities, which are crucial for supporting the massive scaling required by the global energy transition. Our results also serve as a direct example of a near-term quantum application in anomaly detection, consolidating a framework that enables viable solutions with restricted computing resources. We can then deduce that quantum classification can benefit from classical learned preprocessing of the data, further improved by the shared-network structure between the compression and classification networks, which allows the hybrid model to modify the learned representation, shifting emphasis from the reconstruction task toward the classification objective.

\bibliography{sample}

\section*{Author Contributions}
E.C. carried out the algorithm implementation and experiments. E.C. and F.C. conceived the methodology, while A.A. and F.C. proposed and supervised the project. All authors contributed to the discussion, analysis of the results and the writing of the manuscript.
\section*{Additional information}
\subsection*{Data Availability}
The dataset used in this work can be accessed at \url{https://github.com/clayton-h-costa/pv_fault_dataset}. The underlying code for this study is not publicly available but may be made available to qualified researchers on reasonable request from the corresponding author.

\subsection*{Competing Interests}
The authors declare no competing interests.

\subsection*{Funding}
This study received no funding.

\end{document}